\documentclass[12pt,letterpaper]{article}
\pdfoutput=1
\usepackage{graphicx,array}
\usepackage{color}
\usepackage{latexsym}
\usepackage{amsthm}
\usepackage{amsmath}
\usepackage{amssymb}

\usepackage{dsfont}
\usepackage{epsfig}
\usepackage{slashed}
\usepackage{bbold}
\usepackage{psfrag}
\usepackage[svgnames]{xcolor}
\PassOptionsToPackage{caption=false}{subfig}
\usepackage{subcaption}
\usepackage{xfrac}
\usepackage{multirow}
\usepackage{booktabs}
\usepackage[colorlinks=true,linkcolor=MediumBlue,citecolor=Green,urlcolor=violet]{hyperref}
\usepackage{cite}
\usepackage[normalem]{ulem}

\newcommand{\be}{\begin{equation}}
\newcommand{\ee}{\end{equation}}
\newcommand{\bea}{\begin{eqnarray}}
\newcommand{\eea}{\end{eqnarray}}

\def\({\left(}
\def\){\right)}

\usepackage{soul}

\begin{document}
\begin{flushright}
EFI-16-25
\end{flushright}
\vspace{.6cm}
\begin{center}
{\LARGE \bf The photo-philic QCD axion}
\bigskip\vspace{1cm}{

Marco Farina$^1$, Duccio Pappadopulo$^2$, Fabrizio Rompineve$^{2,3}$, and Andrea Tesi$^4$ }
\\[7mm]
 {\it \small
$^1$New High Energy Theory Center, Department of Physics, Rutgers University,\\
  136 Frelinghuisen Road, Piscataway, NJ 08854, USA \\
$^2$Center for Cosmology and Particle Physics, Department of Physics, \\ 
New York University, New York, NY 10003, USA\\

$^3$ Institute for Theoretical Physics, University of Heidelberg, \\
Philosophenweg 19, 69120 Heidelberg, Germany\\

$^4$Enrico Fermi Institute, Department of Physics, University of Chicago,\\ 
5620 S Ellis Ave, Chicago, IL 60637, USA\\
 }

\end{center}

\bigskip \bigskip \bigskip \bigskip

%%%%%%%%%%%%%%%%%%%%%%%%%%%%%%%%%%%%%%%%%%%%%%%%%%%%%%%%%%%%%%%%%%%%%%%%%%
\centerline{\bf Abstract} 
\begin{quote}
We propose a framework in which the QCD axion has an exponentially large coupling to photons, relying on the ``clockwork'' mechanism. We discuss the impact of present and future axion experiments on the parameter space of the model. In addition to the axion, the model predicts a large number of pseudoscalars which can be light and observable at the LHC. In the most favorable scenario, axion Dark Matter will give a signal in multiple axion detection experiments and the pseudo-scalars will be discovered at the LHC, allowing us to determine most of the parameters of the model.
\end{quote}

%%%%%%%%%%%%%%%%%%%%%%%%%%%%%%%%%%%%%%%%%%%%%%%%%%%%%%%%%%%%%%%%%%%%%%%%%%
\vfill
\noindent\line(1,0){188}
{\scriptsize{ \\ \texttt{$^1$ \href{mailto:farina.phys@gmail.com}{farina.phys@gmail.com}\\ $^2$ \href{mailto:duccio.pappadopulo@gmail.com}{duccio.pappadopulo@gmail.com}\\ $^3$ \href{mailto:rompineve@thphys.uni-heidelberg.de}{f.rompineve@thphys.uni-heidelberg.de}\\ $^4$ \href{mailto:atesi@uchicago.edu}{atesi@uchicago.edu}}}}
\newpage

%%%%%%%%%%%%%%%%%%%%%%%%%%%%%%%%%%%%%%%%%%%%%%%%%%%%%%%%%%%%%%%%%%%%%%%%%%%%%%%%%%%%%%%%%%%%%%%%%%%%%%%%%%%%%%%%%%%%%%%%%%%%%%%%%%%%%%%%%%%%%%%%%%%%%%%%%%%%%%%%%%%%%%%%%%%%%%%%%%%%%%%%%%%%%%%%%%%%%%%%%%%%%%%%%%%%%%%%%%%%%%%%%%%%%%%%%%
%%%%%%%%%%%%%%%%%%%%%%%%%%%%%%%%%%%%%%%%%%%%%%%%%%%%%%%%%%%%%%%%%%%%%%%%%%
\section{Introduction}
\label{sec:intro}
%%%%%%%%%%%%%%%%%%%%%%%%%%%%%%%%%%%%%%%%%%%%%%%%%%%%%%%%%%%%%%%%%%%%%%%%%%

The strong CP problem arises from the experimental observation that CP symmetry is respected  by strong interactions with very high precision. In particular, bounds on the neutron electric dipole moment (nEDM) at the level of $d_n< 3\times 10^{-26}\,e$\,cm \cite{EDMn}, require the QCD theta angle to be tiny $\theta_{\rm QCD}\lesssim10^{-10}$.  
A very elegant solution to this puzzle is embodied by the QCD axion~\cite{assionePQ,weinberg,wilczek}. The axion is a pseudo-Nambu-Goldstone boson with a non vanishing potential generated by strong interactions, which dynamically relaxes the $\theta$-angle to 0. Furthermore, the energy density stored in the oscillations of the field around the minimum of its potential constitute, in an expanding universe, a viable cold Dark Matter (DM) candidate. For all these reasons, the QCD axion represents a minimal and compelling extension of the Standard Model (SM).

Many experiments have been built to detect axion DM, with many others that will be operating in the near future. Interestingly, most of them rely on the existence of a coupling between the axion field $a$ and the photons, described by the interaction lagrangian
\be\label{aFF}
-\frac{g_{a\gamma\gamma}}{4} a F_{\mu\nu}\tilde F^{\mu\nu}\,.
\ee
The coupling $g_{a\gamma\gamma}$ is generally non vanishing for the QCD axion, and typical models display a strong correlation between the axion mass and the coupling to photons
\be\label{axionwindow}
\frac{g_{a\gamma\gamma}}{10^{-16}\,{\textrm{GeV}^{-1}}}\sim \frac{m_a}{10^{-6}\,{\textrm{eV}}}~.
\ee
Currently, only a very small fraction of the QCD axion DM window is being probed experimentally by the ADMX experiment~\cite{admx}. Existent proposals for future experiments, like improvement of the ADMX setup \cite{admxfuture} or new axion detection techniques like ABRACADABRA \cite{abra} are expected to increase the sensitivity on the parameter space defined by Eq.~(\ref{axionwindow}), but they are still only able to cover a fraction of the whole QCD axion DM window. Most of the constraining power of axion DM detection experiments reside in the large $g_{a\gamma\gamma}$ region, outside of the range roughly defined by Eq.~(\ref{axionwindow}).

It is however important to keep in mind that the strong correlation between $m_a$ and $g_{a\gamma\gamma}$ described by Eq.~(\ref{axionwindow}) is not a generic consequence of solving the strong CP problem with an axion. The purpose of this paper is to show that the coupling $g_{a\gamma\gamma}$ can indeed be arbitrarily large without spoiling CP conservation by strong interactions. We will show, for the first time, how this is indeed possible using a special realization of the clockwork mechanism introduced by~\cite{choi,rattazzi} (see also \cite{giudice} for a broad overview, and \cite{riotto} for an application to inflation). 

The existence of models in which the QCD axion couples to photons with arbitrary strength broadens the scope of experiments like ADMX and ABRACADABRA, and opens up region in the axion parameter space covered by more than one experiments leading to a rich phenomenology with multiple signals and possible collider smoking guns for our construction.\\

The structure of the paper is as follows. In Section \ref{sec:review} we review the axion solution of the strong CP problem, its viability as a cold Dark Matter (DM) candidate, and the existing and planned experiments to detect it. As the majority of these experiments relies on the coupling of the axion to photons, in Section \ref{sec:eft} we review the predictions of the simplest QCD axion models showing that they imply a very strong correlation between this coupling and the axion mass. In Section \ref{sec:CW} we review the `clockwork mechanism' which allows to obtain exponentially large axion decay constants and we show how this can be used to get arbitrarily large couplings of the axion to photons without spoiling its solution to the strong CP problem nor its validity as a DM candidate. In section \ref{sec:pheno} we discuss the phenomenological implication of this construction. We conclude in section \ref{sec:conclusions}.

%%%%%%%%%%%%%%%%%%%%%%%%%%%%%%%%%%%%%%%%%%%%%%%%%%%%%%%%%%%%%%%%%%%%%%%%%%%%%%%%%%%%%%%%%%%%%%%%%%%%%%%%%%%%%%%%%%%%%%%%%%%%%%%%%%%%%%%%%%%%%%%%%%%%%%%%%%%%%%%%%%%%%%%%%%%%%%%%%%%%%%%%%%%%%%%%%%%%%%%%%%%%%%%%%%%%%%%%%%%%%%%%%%%%%%%%%%%%%%%%%%%%%%%%%%%%%%%%%%%%%%%%%%%%%%%%%%%%%%%%%%%%%%%%%%%%%%%%%%%%%%%%%%%%
\section{A lightning review of axion physics}
\label{sec:review}
%%%%%%%%%%%%%%%%%%%%%%%%%%%%%%%%%%%%%%%%%%%%%%%%%%%%%%%%%%%%%%%%%%%%%%%%%%
The axion is a Nambu-Goldstone boson of the Peccei-Quinn (PQ) symmetry \cite{assionePQ} and the strong CP problem is solved by the following interaction \cite{assionePQ,weinberg,wilczek}
\be\label{CP1}
\mathcal L_a\supset-\frac{\alpha_s}{8\pi}\left(\frac{a}{f_a}-\theta_{\rm QCD}\right)G_{\mu\nu}^A\tilde G^{A\mu\nu}.
\ee
Non perturbative QCD effects generate a potential for $\bar a\equiv a-f_a\theta_{\rm QCD}$, that relaxes the effective vacuum angle $\langle\bar a\rangle$ to zero, enforcing CP conservation.\footnote{We use $\tilde G_{\mu\nu}^A=1/2 \epsilon_{\mu\nu\alpha\beta}G^{A\alpha\beta}$.} In particular the axion receives a small mass
\be\label{mass}
m_a\approx 5.7\times 10^{-10}\,{\textrm{eV}}\left(\frac{10^{16}\,{\textrm{GeV}}}{f_a}\right).
\ee

In models in which the axion field arises from the spontaneous breakdown of a global U(1) symmetry \cite{KSVZ1,KSVZ2, DFSZ1,DFSZ2} the effective interactions in Eq.~(\ref{CP1}) is generated if the theory contains colored fields carrying U(1) charge and if this symmetry has a color anomaly. Depending on the other gauge quantum numbers of the colored fermions (that can be the SM fermions in certain realizations \cite{DFSZ1,DFSZ2}), the axion Lagrangian will contain additional interactions. Above the scale of QCD confinement (and the weak scale), the most general lagrangian describing axion interaction can be written as
\be \label{axionUV}
\begin{split}
\mathcal L_a&=\frac{1}{2}(\partial_\mu a)^2-\frac{g_s^2}{32\pi^2} \frac{a}{f_a}\, G^{A\mu\nu}\tilde G_{\mu\nu}^A-c_B\frac{g'^2}{32\pi^2} \frac{a}{f_a}\, B^{\mu\nu}\tilde B_{\mu\nu}-c_W\frac{g^2}{32\pi^2} \frac{a}{f_a}\, W^{a\mu\nu}\tilde W^a_{\mu\nu}\\
&+  \frac{\partial_\mu a}{2f_a} \sum_i c_i \bar \psi_i\gamma^\mu\gamma_5\psi_i  + c_H \frac{\partial_\mu a}{2f_a} \left( i H^\dag D_\mu H - i (D_\mu H)^\dag H\right)+O(1/f_a^2),
\end{split}
\ee
where the sum runs over all the SM fermions.

%%%%%%%%%%%%%%%%%%%%
%%%%%%%%%%%%%%%%%%%%
%%%%%%%%%%%%%%%%%%%%
\begin{figure*}[tb]
\centering
\includegraphics[width=0.65\linewidth]{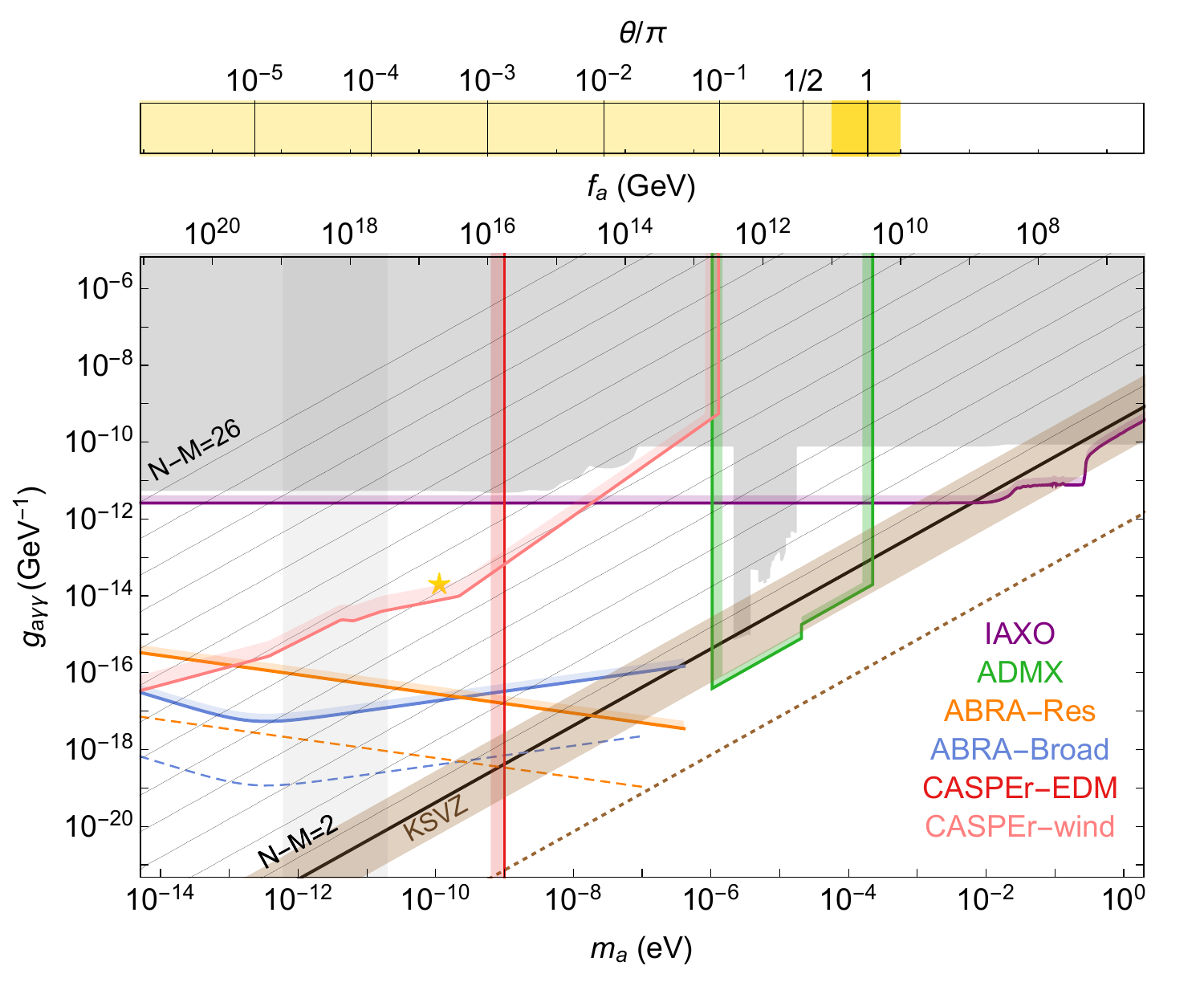}
\caption{{\small\it The axion parameter space as a function of its mass $m_a$ (and corresponding decay constant $f_a$) and its coupling to photons $g_{a\gamma\gamma}=\frac{c_\gamma}{f_a}\frac{\alpha}{2\pi}$. The colored lines represent the projected sensitivities of future experiments, namely ABRACADABRA~\cite{abra}, ADMX~\cite{admxfuture}, phase II CASPEr-EDM~\cite{casper} and IAXO~\cite{iaxo}. 
We also show the bound from the CASPEr-wind experiment proposed in \cite{Graham:2013gfa}, for a $^3$He target and the model in Sec.~\ref{sec:CW}.
In gray the region already excluded either by astrophysical observations or experiments as described in the text. The black solid line and brown band correspond to possible KSVZ QCD axion models (see section \ref{sec:KSVZ}).  The thin lines illustrate the prediction of the clockwork QCD axion for $N-M = 2k$ with $k=1,..,13$ and with fermionic content as described in section \ref{sec:pheno}. The star indicates the specific example used in Fig.~\ref{figexample}. In the upper band we show the initial misalignment angle needed to reproduce the observed DM relic density. The darker yellow band corresponds to the uncertainty on $f$ for the axion to be DM in the Pre-inflationary axion scenario.}}
\label{plottone}
\end{figure*}
%%%%%%%%%%%%%%%%%%%%
%%%%%%%%%%%%%%%%%%%%
%%%%%%%%%%%%%%%%%%%%

Below the QCD confinement scale a potential for the axion is generated by strong interactions. Furthermore the couplings in Eq.~(\ref{axionUV}) generate interactions of the axion with photons, neutrons, protons, and electrons
\be \label{effaxion}
\mathcal L_a=\frac{1}{2}(\partial_\mu a)^2 - V(\bar a)-c_\gamma\frac{a}{f_a}\frac{\alpha}{8\pi} \, F^{\mu\nu}\tilde F_{\mu\nu}+  \frac{\partial_\mu a}{2f_a} \left(c_n \bar{n}\gamma^\mu\gamma_5 n + c_p \bar{p}\gamma^\mu \gamma_5 p + c_e \bar{e}\gamma^\mu\gamma_5 e\right).
\ee
The couplings appearing in \eqref{effaxion} depend both on the fundamental parameters of \eqref{axionUV} and on QCD infra-red effects. The couplings to photons, neutrons, protons, and electrons are present at low energy even if they are absent in the UV. The axion has its minimum for $\langle a \rangle = f_a \theta_{\rm QCD}$ and $V(\bar a)$ is a cosine function, approximately \cite{preciso}.

The very small axion mass in Eq.~(\ref{mass}) makes the axion a viable DM candidate. As the universe cools down and the Hubble parameter $H$ becomes smaller than $m_a$, the axion field starts to oscillate coherently around the minimum of its potential. The energy density stored in these oscillations behave as Cold Dark Matter (CDM). This fixes the amplitude of axion oscillations today to be
\be\label{DMamplitude}
\theta_{\textrm{today}}\equiv\frac{a_{\textrm{today}}}{f_a}\approx 3.6\times10^{-19}.
\ee

The initial conditions for the axion field leading to Eq.~(\ref{DMamplitude}) and the explicit expression of the axion abundance in terms of those depend on the cosmological history \cite{D'Eramo:2014rna}. There are two well known regimes depending on the relative size of the PQ symmetry breaking scale $\Lambda_{\rm PQ}\sim f_a$, the Hawking temperature during inflation $T_{\rm H}=H_I/2\pi$ and the maximal temperature achieved by the universe during reheating, $T_{\rm max}$.\footnote{This temperature coincide with the standard reheating temperature for instantaneous reheat, but can in principle be much larger \cite{D'Eramo:2014rna}.}
\begin{itemize}
\item Pre-inflationary axion, $\Lambda_{\rm PQ}>\max(T_{\rm H},T_{\rm max})$: the PQ symmetry is already broken during inflation and does not get restored by reheating. Then the initial misalignment angle $\theta$ is a fundamental constant in our Hubble patch and the axion energy density is given by
\be\label{axionpre}
\Omega_a^{\textrm{Pre}}h^2\approx0.2\left(\frac{f_a}{10^{12}\,{\textrm{GeV}}}\right)^{1.18}\theta^2 F(\theta),
\ee
where $F(\theta)$ accounts for anharmonicities of the axion potential and $F(0)=1$. Eq.~(\ref{axionpre}) is valid for $f_a\lesssim 10^{16}$\,GeV, when the axion starts to oscillate before the QCD phase transition. This scenario is constrained by isocurvature perturbations.
\item Post-inflationary axion, $\Lambda_{\rm PQ}<\max(T_{\rm H},T_{\rm max})$: the PQ symmetry in not broken during inflation or gets restored during reheating. In this case Eq.~(\ref{axionpre}) has to be averaged over $\theta$ and the absence of stable domain walls is required. An additional and uncertain contribution to $\rho_a$ in the Post-inflationary case comes from string evolution and domain walls collapse resulting in a possible range of $f_a$ given by
\be
f_a^{\textrm{Post}}=(10^{11}-10^{12})\,{\textrm{GeV}}.
\ee
\end{itemize}

Detection of axion DM is a very active experimental field. Different axion DM detection experiments exploits different couplings appearing in Eq.~(\ref{effaxion}). The only model independent prediction of QCD axion DM is the existence of the coupling in Eq.~(\ref{CP1}) above the QCD confinement scale. This coupling induces a time varying nuclear EDM that can be looked for experimentally. This is the idea behind the CASPEr-EDM experiment \cite{casper} which plans to cover, in its Phase 2 stage, the relevant region for QCD axion DM for $m_a\lesssim 10^{-9}$\,eV, or equivalently $f_a\gtrsim 10^{16}$\,GeV. A similar experimental setup, CASPER-wind  \cite{Graham:2013gfa}, is also able to probe the axial couplings of the axion to quarks.

Currently, the strongest constraints on QCD axion DM, even though limited to a narrow mass region around $m_a\approx 10^{-6}$\,eV, come from the ADMX experiment~\cite{admx}. ADMX exploits the fact that if the coupling $c_\gamma$ is non vanishing below the QCD phase transition (which is generically the case for the QCD axion), then in the presence of a constant magnetic field, axion DM can generate an electric field oscillating with frequency $\omega=m_a c^2/\hbar$, the axion Compton frequency. This oscillating electric field can in turn be detected by a cavity whose size is comparable to the axion Compton wavelength. This resonant effect turns out to be a limitation for ADMX to extend its reach to smaller axion masses (larger values of $f_a$), as they would require very large cavities. However future upgrades of the experiment~\cite{admxfuture}, including an High Frequency run, should cover more than 2 decades in axion mass.

A recent development in the detection of axion DM, in particular through the coupling to photons, comes from the  ABRACADABRA experiment~\cite{abra}. In this experiment an oscillating magnetic field, generated by the axion field oscillations, is detected through a superconducting SQUID loop. This approach does not rely on the observation of a cavity resonance and can in principle cover a large range of axion masses. 

The planned reach of ADMX-HF, ABRACADABRA, the second phase of operation of CASPEr-EDM and CASPER-wind, are all shown in Fig.~\ref{plottone} in the $(m_a, \,g_{a\gamma\gamma})$ plane under the assumption that the axion constitutes the whole DM energy density.\footnote{Since CASPEr exclusion only depends on the axion mass and its coupling to nucleons, it is a vertical line in the $(m_a, \,g_{a\gamma\gamma})$ plane.} We also show the reach on the parameter space of the model presented in Sec.~\ref{sec:CW}, of the experiment proposed in \cite{Graham:2013gfa}, aiming to detect axion couplings to nuclear axial currents.
As the figure shows, these experiments do not only probe the QCD DM axion line but they cover a much wider fraction of the parameter space in which the coupling to photons can be several orders of magnitude larger than the standard QCD axion prediction (represented by the brown shaded region in Fig~\ref{plottone}).

The gray region in Fig.~\ref{plottone} represents a collection of existing bounds on the axion parameter space both from experiments and from astrophysical observations. Apart from the present ADMX exclusion the gray region contains bounds from helioscopes (CAST)~\cite{CAST}, from the lack of detection of a $\gamma$ ray burst associated with supernova SN1987A~\cite{supernova}, non observation of spectral irregularities by Fermi-LAT ~\cite{fermi} and H.E.S.S.~\cite{hess},  and from astrophysical observations related to horizontal branch stars in Galactic Globular Clusters~\cite{hbstars}. The vertical gray band corresponds to the region excluded by black hole superrandiance~\cite{superradiance}. 
For a more detailed discussion of the bounds we refer the reader to \cite{PDG} and references therein.  
Finally Fig.~\ref{plottone} includes projections from next generation helioscopes, IAXO~\cite{iaxo}.

%%%%%%%%%%%%%%%%%%%%%%%%%%%%%%%%%%%%%%%%%%%%%%%%%%%%%%%%%%%%%%%%%%%%%%%%%%
\section{The standard QCD axion}
\label{sec:eft}
%%%%%%%%%%%%%%%%%%%%%%%%%%%%%%%%%%%%%%%%%%%%%%%%%%%%%%%%%%%%%%%%%%%%%%%%%%
In this section we review how KSVZ models solve the strong CP problem and we discuss the size of the coupling of the axion to photon (and matter) they predict.\footnote{For a similar discussion of DFSZ models see \cite{PDG}.}
%%%%%%%%%%%%%%
\subsection{KSVZ model}\label{sec:KSVZ}
%%%%%%%%%%%%%%
%One introduces a complex scalar field $\phi$ and a set of fermion $\psi_r$, $\psi_{\bar r}$ transforming under representations $r$ and $\bar r$ of the SM gauge group. 

In the original KSVZ model \cite{KSVZ1,KSVZ2} the field content is a complex scalar field $\phi$ and a vector-like pair of color triplet fermions $\psi$, $\psi^c$. The action of the PQ symmetry is $\phi\to e^{i\alpha}\phi,\quad \psi\to e^{-i\alpha}\psi$, and therefore forbids a Dirac mass for the fermions. The lagrangian of the model is
\be
\mathcal{L}_{\rm KSVZ}\supset \mathcal{L}_{\rm SM} +   |\partial_\mu \phi|^2 - V(|\phi|^2) - y (\phi \psi \psi^c +{\textrm{h.c.}}) + \cdots
\ee
The U(1) symmetry is assumed to be broken spontaneously with $\phi = (v + \varphi/\sqrt 2) \exp(i a/\sqrt 2v)$, where $\varphi$ is a radial mode and $a$ the axion. The fermions obtain a mass $m_\psi=y v$. Integrating out the colored fermions assuming $yv\gg m_Z$, we get eq. \eqref{axionUV} where however all the coefficients are vanishing and $f_a \equiv \sqrt{2}v$. Upon confinement the axion $a$ inherits couplings to photons and nucleons \cite{preciso}
\be\label{KSVZmathcing}
c_\gamma = -\frac{2}{3}\frac{4m_d+m_u}{m_d+m_u}\simeq -1.92,\quad c_p\simeq -0.47 ,\quad c_n\simeq -0.02.
\ee
These contributions are model independent effects from low energy QCD. The prediction of the standard KSVZ axion is shown as a thick black line in Fig.~\ref{plottone}.

%%%%%%%%%%%%%%
\subsection{Variations over KSVZ}
%%%%%%%%%%%%%%
Many variations can be built starting from the simplest model. In particular in the case where PQ acts on a new sector of fields, we can explore the presence of more than one vector-like pairs. The KSVZ can be easily extended to allow for generic representations
\be
\mathcal{L} \supset |\partial_\mu \phi|^2 - V(|\phi|^2) - \sum_r y_r \phi \psi_r \psi_{r^c} +\cdots,
\ee
where $r=(r_3,r_2,r_Y)$ is an irreducible representation of SU(3)$_c\times$ SU(2)$\times$U(1)$_Y$, and $r^c$ its conjugate. After the spontaneous PQ breaking, $\langle \phi\rangle=v$, we can integrate out all the heavy fermions and we generate the following interactions,
\be\label{coup-ksvz}
\frac{1}{f_a}\equiv\frac{|\sum_r 2 \mathrm{dim}(r_2) T(r_3) |}{\sqrt{2}v}, \quad c_W = \frac{\sum_r 2 \mathrm{dim}(r_3) T(r_2) }{|\sum_r 2 \mathrm{dim}(r_2) T(r_3)| }, \quad c_B = \frac{\sum_r 2 \mathrm{dim}(r_3)\mathrm{dim}(r_2) T(r_Y) }{|\sum_r 2 \mathrm{dim}(r_2) T(r_3) |},
\ee
where $T(r)$ is the index of the representation $r$ (it is the charge squared for U(1)$_Y$) and $\dim(r)$ its dimension. In this model we define the Domain Wall (DW) number,
\be
N_{\rm DW}=\left|\sum_r 2 \mathrm{dim}(r_2) T(r_3) \right|
\ee
and 
\be
E=\sum_r 2 \mathrm{dim}(r_3) T(r_2) +\sum_r 2 \mathrm{dim}(r_3)\mathrm{dim}(r_2) T(r_Y) =\sum_r 2 \mathrm{dim}(r_3) Q^2_r,
\ee
where $Q_r$ is the electric charge.
The low energy couplings are,
\be
c_\gamma \approx \frac{E}{N_{\rm DW}} - 1.92,\quad c_p \simeq -0.47,\quad c_n\simeq -0.02.
\ee
Using different representations it is possible to deviate from the standard KSVZ scenario, especially below the QCD line in the $(m_a, c_\gamma)$ plane. As discussed in \cite{diluzio} when two or more representations are present it is conceivable to get a cancellation in the predicted value of $c_\gamma$. The brown band in Fig.~\ref{plottone} shows the prediction of the set of KSVZ models listed in Table II of \cite{diluzio}. They all use a single irreducible representation of the SM gauge group to generate both the color and electromagnetic anomaly. The dashed brown line in the same plot shows that when more than one irrep is used cancellation may occur, as it is the case if the fermionic content is for instance $r_1=(3,3,-1/3)$ and $r_2=(\bar 6, 1,-1/3)$, and $E/N=23/12\approx1.917$.
\\

We want to emphasize, however, that is difficult to robustly amplify $c_\gamma$ in this context, since the models quickly require very large hypercharges and/or large SU(2) representations.
In the following section we describe a model that easily enhances $c_\gamma$ and naturally predicts values for the axion coupling to photons that will be testable at future experiments.

%%%%%%%%%%%%%%%%%%%%%%%%%%%%%%%%%%%%%%%%%%%%%%%%%%%%%%%%%%%%%%%%%%%%%%%%%%
\section{The clockwork QCD axion}
\label{sec:CW}
%%%%%%%%%%%%%%%%%%%%%%%%%%%%%%%%%%%%%%%%%%%%%%%%%%%%%%%%%%%%%%%%%%%%%%%%%%
In this section we review the clockwork mechanism \cite{rattazzi} and show how it can be used to generate a QCD axion with an exponentially large coupling $c_\gamma$.\footnote{The observation the a clockwork axion can solve the strong CP problem is also found in \cite{fuminobu}. $c_\gamma$ is however $O(1)$ in that paper.}

\subsection{The clockwork mechanism}
Consider a set of $N+1$ complex scalar fields $\phi_n$, $n=0,\ldots,N$, with the following renormalizable potential
\be\label{CW}
V(\phi)=\sum_{n=0}^N\left(-m_n^2|\phi_n|^2+\lambda_n|\phi_n|^4\right)+\sum_{n=0}^{N-1}\left(\kappa_n\phi^\dagger_n\phi^3_{n+1}+{\textrm{h.c.}}\right)
\ee
The potential in Eq.~(\ref{CW}) has an unbroken U(1)$_{\rm CW}$ symmetry under which $\phi_n$ has charge $1/3^{n}$. In the limit in which $\kappa_n=0$ a full U(1)$^{N+1}$ global symmetry is restored. A whole set of U(1)$^{N+1}$ symmetric quartic couplings $|\phi_n|^2|\phi_m|^2$ has been neglected for simplicity. To simplify the discussion we will assume $m_0^2=\ldots=m_N^2\equiv m^2$, $\lambda_0=\ldots=\lambda_N\equiv\lambda$ and $\kappa_0=\ldots=\kappa_N\equiv\kappa$. Furthermore it is natural to assume $\kappa\ll\lambda$ as this condition is not spoiled by renormalization. In this limit, the scalar fields can be expanded around their vacuum expectation value
\be\label{phin}
\phi_n=\left(v+\frac{\varphi_n}{\sqrt 2}\right)e^{\frac{i\pi_n}{\sqrt 2 v}}, \quad v=\frac{ m}{\sqrt2\lambda}.
\ee
The mass of the radial modes $\varphi_n$ is of order $\sqrt\lambda v$, so that in the limit $\lambda\gg\kappa$ they are much heavier than the angular modes. We will thus neglect the radial variables and focus on the angular ones. Their spectrum contains a massless mode, the Nambu-Goldstone mode from the breakdown of U(1)$_{\rm CW}$ and a set of $N$ massive modes with masses of order $\sqrt\kappa v$.

Plugging Eq.~(\ref{phin}) back into Eq.~(\ref{CW}) one obtains the following mass term for the $\pi$s
\be\label{masstermCW}
V(\pi)\supset \frac{\kappa v^2}{2}\sum_{n=0}^{N-1}(\pi_n-q \pi_{n+1})^2
\ee
where in our case $q=3$. The mass matrix for the angular mode is thus tridiagonal due to the nearest neighbors nature of the quartic interactions.
%\be\label{MCW}
%M^2=\kappa V^2\begin{pmatrix}
%1 & -q & 0 & 0 & &.  &. &.  \\
%-q & 1+q^2 & -q & 0 &  &  &\\
%0 &-q & 1+q^2 & -q &  &  &\\
%. &  &   &   &. & &\\
%  &  &   &   &  & . &\\
%. &  &   &   & & &  &\\
% &  &   &   &  0& -q& 1+q^2 &-q\\
%. &  &   &   &0  & 0 & -q & q^2\\
%\end{pmatrix}
%\ee
Writing a generic eigenvector of the matrix $M^2$ as $u=(a_0,\ldots, a_N)^T$, such that $M^2\cdot u=\lambda u$ and introducing two fictitious entries $a_{-1}$ and $a_{N+1}$, one has the following set of constrains
\begin{align}\nonumber
 a_{-1}-q  a_0&=0\\
 a_{N}-q  a_{N+1}&=0\\\nonumber
 -qa_{n-1}+(1+q^2)a_n-q a_{n+1}&=\lambda a_n\quad (n=1,..., N-1)
\end{align}
The homogeneous difference equation on sites $1$ to $N-1$ is formally solved by the ansatz $a_n=u w^n$, with
\be\label{chareq}
w^2+\left[\frac{\lambda-(1+q^2)}{q}\right]w+1=0.
\ee
Eq.~(\ref{chareq}) is the characteristic equation associated to the linear recurrence. A generic solution of the recurrence is given by $u_1 w_1^n+u_2 w_2^n$. Notice that $w_{1,2}$ are such that $w_1 w_2=1$.  This implies that if $w_1$ and $w_2$ are complex $w_{1,2}=e^{\pm i\theta}$, with real $\theta$.

%%%%%%%%%%%%%%%%%%%%%%
\begin{itemize}
\item Real $w_{1,2}$. Let us write $w_1=w$ and $w_2=1/w$, with $w>1$. Let us also assume that $u_{1}\neq 0$ and set $u_1=1$ and $u_2=u$. One can then show that the boundary conditions at the zeroth and $N$th site are never satisfied. Allowing $u_1=0$ on the other hand both boundary conditions imply $w=1/q$ which in turn gives $\lambda=0$. This solution defines the massless mode for which
\be\label{a0}
\lambda^{(0)}=0,\quad a_n^{(0)}=\frac{C}{q^n},\quad C=\sqrt{\frac{1-q^{-2}}{1-q^{-2(N+1)}}}.
\ee 
%%%%%%%%%%%%%%%%%%%%%%
\item Complex $w_{1,2}$. The characteristic equation has complex solutions for $(1-q)^2<\lambda<(1+q)^2$, so that the clockwork spectrum will be bounded between these two values. Furthermore we can rewrite the solution of the recursion as
\be\label{sol}
a_n=\sin n\theta+ u\cos n\theta.
\ee
Plugging this back in the recursion relation requires
\be
\lambda=1+q^2-2q\cos\theta.
\ee
The boundary conditions at sites $0$ and $N$ imply
\begin{equation}\label{BC}
u=\frac{\sin\theta}{\cos\theta-q}~~~{\textrm{and}}~~~ -q\sin N\theta+(1+q^2)\sin(N+1)\theta-q\sin (N+2)\theta=0.
\end{equation}
The second boundary condition is solved for all $\theta$ of the form
\be
\theta_i=\frac{i\pi}{N+1},
\ee
with integer $i$. Given Eq.~(\ref{sol}) and (\ref{BC}) the only valid solutions are for $i=1,\ldots, N$. These provide the $N$ eigenvectors with non-vanishing eigenvalues
\be\label{massive}
\lambda^{(i)}=1+q^2-2q\cos\frac{i\pi}{N+1}, \quad a_n^{(i)}=\sqrt{\frac{2}{(N+1)\lambda^{(i)}}}\left(\sin\frac{(n+1)i\pi}{N+1}-q\sin\frac{ni\pi}{N+1}\right)
\ee
\end{itemize}
%%%%%%%%%%%%%%%%%%%%
Combining eq.s~\eqref{a0}-\eqref{massive} we can now define the mass eigenstates as $A_i$ with masses $m_i$ to be
\be\label{rotation}
A_i=\sum_{n=0}^N a^{(i)}_n \pi_n\quad{\textrm{or}}\quad \pi_i= \sum_{n=0}^NA_n a^{(n)}_i
\ee
\be\label{masses}
m_0=0,\quad m_i=\left(1+q^2-2q\cos\frac{i\pi}{N+1}\right)^{1/2}\sqrt{\kappa}v. 
\ee
Notice that for fixed $N$ the mixing angles are bounded by $\sqrt{2/(N+1)}$ as a consequence of having a large number $N$ of states.
%where we used the fact the the matrix $R_{ij}\equiv a^{(i)}_j$ is an orthogonal $N+1$ by $N+1$ matrix. 
The spontaneously broken U(1)$_{\rm CW}$ symmetry under which $\phi_n\rightarrow e^{i Q_n\alpha}\phi_n$ with $Q_n=\frac{Q_0}{q^n}$ is non linearly realized on the angular fields as
\begin{equation}\label{nonlinear}
a\equiv A_0\to a+\frac{\sqrt 2 v}{C} Q_0\alpha.
\end{equation}
while $A_n\to A_n$ for $n=1,\ldots,N$. $Q_0$ is an arbitrary normalization of the U(1)$_{\rm CW}$ charge that can be set to one, and $C \stackrel{N\gg1}{\approx} 2 \sqrt{2}/3\approx 0.94$. Eq.~(\ref{nonlinear}) defines the canonical axion decay constant to be
\be\label{decayf}
f \equiv \frac{\sqrt{2} v}{C}.
\ee

\begin{figure*}[tb]
\centering
\includegraphics[width=1.0\linewidth]{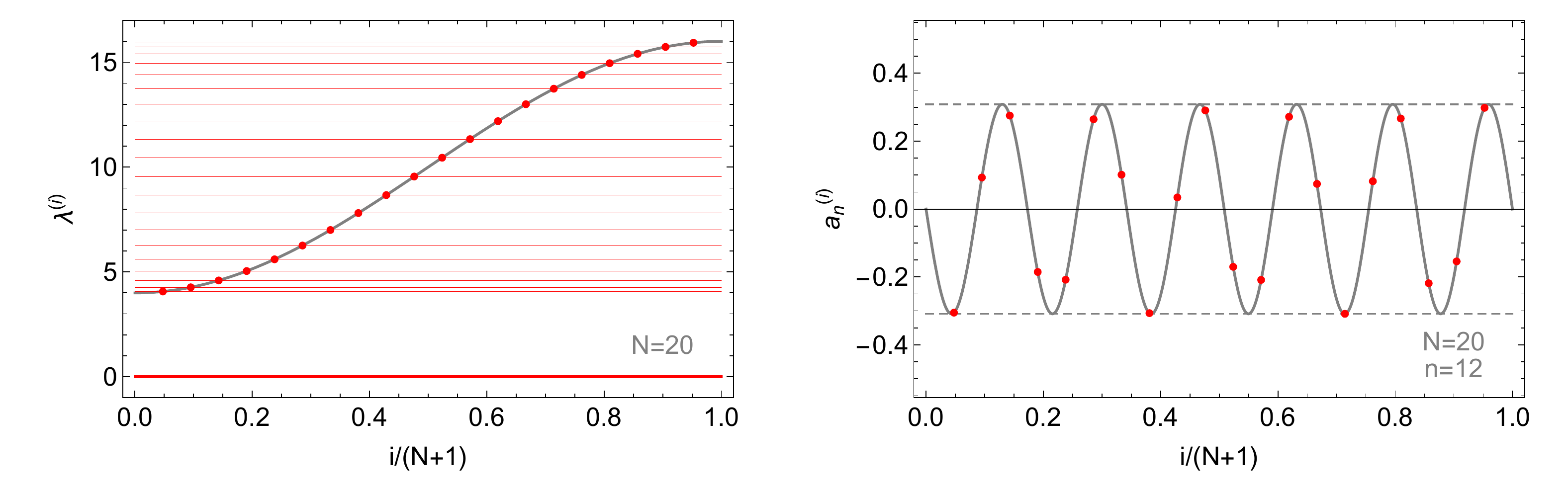}
\caption{{\small\it 
{\bf Left:} the clockwork spectrum for $q=3$ and $N=20$. For different $N$ the massive levels are uniformly spaced on the horizontal axis and lie on the blue curve. {\bf Right:} size of the mixing angle of the massive state $A_i$ with site $n$, for fixed $n=12$ and $N=20$.  Notice that the absolute value of the mixing angle is bounded by $\sqrt{2/(N+1)}$.}}
\label{spectrum}
\end{figure*}

The clockwork spectrum is shown in Fig.~(\ref{spectrum}). The striking feature of this construction is the structure of the wavefunction of the zero mode in field space. The overlap of the massless axion with the high-$n$ sites is exponentially suppressed by a factor $1/q^n$. This implies that while the fundamental NGB decay constant is $f$, the effective decay constant observed by the $n$-th site is exponentially larger, $q^n f$.
%%%%%%%%%%%%%%%%%%%%
%%%%%%%%%%%%%%%%%%%%
%%%%%%%%%%%%%%%%%%%%
%%%%%%%%%%%%%%%%%%%%
%%%%%%%%%%%%%%%%%%%%
%%%%%%%%%%%%%%%%%%%%

\subsection{A photo-philic QCD axion}
Turning the previous construction to a solution to the strong CP problem is easy. It is enough to introduce a pair of colored fermions $\psi_r$ and $\psi_{r^c}$ in a representation $r$, such that  the Yukawa coupling
\be\label{qcdfermions}
y_j\,\phi_j \psi_r\psi_{r^c} 
\ee
is allowed by their U(1)$_{\rm CW}$ charges. Upon the field redefinition in \eqref{rotation}, the original variables $\phi_j$ can be expressed as 
\be
\phi_j \propto v \exp\left(\frac{i\,a}{q^j f}\right), \quad n=0,...,N.
\ee
This makes manifest that the field range for $a$ is $\Delta a = 2\pi q^N f$. When colored vector-like fermions are on the $j$-th site (in general $j\leq N$) as in Eq.~\eqref{qcdfermions}, the effective coupling of the axion to gluons obtained integrating out the colored fermions is as in Eq.~(\ref{CP1}) with
\be\label{faCW}
f_a = \frac{q^j f}{2T(\psi_r)},
\ee
where with $T(\psi_r)$ we refer to the index of the SU(3)$_c$ representation of the fermion $\psi$.
Therefore the DW number of the theory, or alternatively the number of zero energy minima in the axion field range is
\be
N_{\rm DW} = \frac{q^Nf}{f_a} =\frac{2 T(\psi_r) q^N f}{q^j f}= q^{N-j}\, 2 T(\psi_r).
\ee
 In order to have $N_{\rm DW}=1$ we need $j=N$ and just one pair of vector-like fermions in the fundamental of SU(3)$_c$. In order to make this colored fermion unstable we consider the case where $\psi_r$ has the same quantum numbers of the left-handed SM quarks $U$ or $D$, we define it as $\psi_{U,D}$ (and $\psi_{U^c,D^c}$ their vector-like partners). Away from the limit $j=N$, the DW number grows generically as $N_{\rm DW}\approx q^N$. If there is more than one site with colored fermions, the above formula needs to be corrected. \\
  
%%%%%%%%%%%%%%%%%%%%
%%%%%%%%%%%%%%%%%%%%
%%%%%%%%%%%%%%%%%%%%
\begin{figure*}[tb]
\centering
\includegraphics[width=0.6\linewidth]{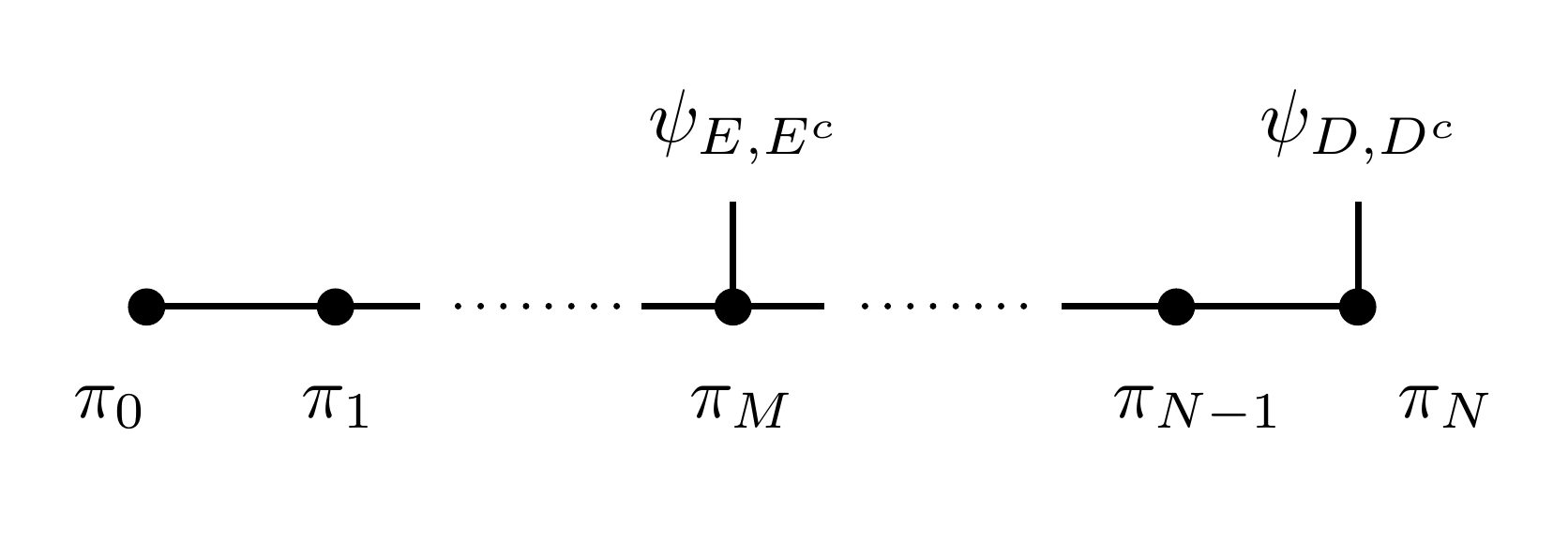}
\caption{{\small\it 
A schematic representation of our construction. Each site represents a scalar field and each link represents the nearest-neighbor structure of the $\pi$ mass terms in Eq.~(\ref{masstermCW}). At sites $N$ and $M$ we couple the vector-like fermions which are responsible for the generation of the color and electromagnetic anomaly.}}
\label{schema}
\end{figure*}
%%%%%%%%%%%%%%%%%%%%
%%%%%%%%%%%%%%%%%%%%
%%%%%%%%%%%%%%%%%%%%

In order to generate an exponentially large coupling of $a$ to photons we introduce, for simplicity, a single set of color neutral vector-like fermions coupled at site $M$, $M<N$
\be\label{photon-fermions}
y_M\,\phi_M \psi_E \psi_{E^c},
\ee
where the fermions have hypercharge $Y(\psi_E)=-Y(\psi_{E^c})=1$ and no SU(2) charge. This choice also ensures that the fermion is unstable and we need not to worry about stable charged relics.\footnote{The decay of the new fermions happens since  $\psi_{U^c,D^c,E^c}$ can have a tree-level mixing with the SM fermion $E,U,D$. Notice that this is fully consistent with the PQ symmetry. Their branching ratios and decay rates are model dependent issues that we do not consider here.} 
Notice that the mass of the fermions coupled to the clockwork sites is set by the size of $\langle\phi_n\rangle\sim f$ and not by the effective decay constant $f_a$.

The axion-photon coupling obtained integrating out the heavy fermions is
 \be
 \frac{E}{N}= \zeta q^{N-M},
 \ee
 where $\zeta=Y(\psi_E)^2/T(\psi_{U,D})=2Y(\psi_E)^2$.
 Therefore at low energies,
 \be\label{cgammaCW}
 c_\gamma \approx \zeta q^{N-M} - 1.92.
 \ee
We have thus found that the coupling to photon can be exponentially large, and its size is controlled by the `distance' between the site coupled to colored fermions and the site coupled to only electro-weak charged ones. Given this dependence, it is also manifest that the contribution to $c_\gamma$ is only marginally affected by sites $i$ with $M<i\leq N$: their presence will only add a correction of $O(q^{-|i-M|})$ to the above formula. A schematic representation of our construction is presented in Fig.~\ref{schema}.

%%%%%%%%%%%%%%%%%%%%%%%%%%%%%%%%%%%%%%%%%%%%%%%%%%%%%%%%%%%%%%%%%%%%%%%%%%%%%%%%%%%%%%%%%%%%

\section{Clockwork axion window: constraints and signals}
\label{sec:pheno}

The parameter space of the clockwork QCD axion is vast. Imposing that the couplings in Eq.~(\ref{CW}) are equal is in principle not necessary. Also, the fermion content can vary. In order to keep our discussion as simple as possible we will make various simplifying assumptions that will not hide any qualitatively new effect. 

We will keep the assumption made below Eq.~(\ref{CW}) of all quartic and masses to be equal. We will furthermore stick to the very simple fermionic content of the previous section, with color triplet fermions $\psi_{U}$ or $\psi_D$ to generate the QCD anomaly and hyper-charge $Y(\psi_E)=1$ singlets to generate the large coupling to photons. In particular, in order to avoid any additional complication with a very large DW number, we will couple the colored fermions to the $N$-th site, implying $N_{\rm DW}=1$. 

Axion physics depends on three parameters, the fundamental symmetry breaking scale $f$, the number of sites $N$, and the site $M$ at which the color singlet and electrically charged fermions couple. Using Eq.~(\ref{faCW}) and (\ref{cgammaCW}) we have
\be
f_a=3^N f
\ee
and
\be\label{gagg}
%g_{a\gamma\gamma}\equiv \frac{\alpha}{2\pi f_a}\left(c_\gamma-1.92\right)\approx\frac{\zeta}{3^M f}\frac{\alpha}{2\pi}.
g_{a\gamma\gamma}\equiv \frac{\alpha}{2\pi f_a}c_\gamma\approx\frac{\zeta}{3^M f}\frac{\alpha}{2\pi}.
\ee
For the special fermion content we are considering $\zeta=2$. The lines of constant $N-M$ in the $(m_a, g_{a\gamma\gamma})$ are shown in Fig.~\ref{plottone}, for the special choice of representation made in this section. Notice however that the whole plane can in principle be filled.

One important observation to be made is that the coupling to photons in Eq.~(\ref{gagg}) generates, at one loop, non vanishing values for the couplings $c_i$ in Eq.~(\ref{axionUV}) of the axion to quarks and leptons. While for $c_\gamma=O(1)$ the size of this contributions is negligible, it can be important for $c_\gamma$ as in Eq.~(\ref{cgammaCW}). We find
\be
c_i=\frac{3\alpha^2 }{2\pi}Q_i^2\zeta q^{N-M}\log\frac{m_\varphi}{\Lambda}\, ,
\ee
where $Q_i$ is the electric charge of the fermion $\psi_i$, $m_\varphi$ is the typical mass of a radial mode, and $\Lambda$ is the QCD confinement scale. Such coupling can be detected by the CASPEr-wind experiment proposed in \cite{Graham:2013gfa}. In Fig.~\ref{plottone} we show the reach of this experiment for a (10\,cm)$^3$ $^3$He target with full nuclear polarization. We assume $m_\varphi\sim 1\,$TeV to draw the line.

For the special case $M=0$, the smallest possible value of $f$ and hence the largest possible value of $N$, is set by the experimental upper bound on $g_{a\gamma\gamma}$. For $f_a$ between $10^8$ and $10^{16}$\,GeV, excluding the ADMX window, the bound coming from CAST is roughly constant given by $g_{a\gamma\gamma}\lesssim 10^{-10}\,{\textrm{GeV}}^{-1}$, which gives
\be\label{m0bound}
f\gtrsim 2\times 10^7\,{\textrm{GeV}}\left(\frac{10^{-10}\,{\textrm{GeV}}^{-1}}{g^{\textrm{max}}_{a\gamma\gamma}}\right)
\ee
or equivalently
\be\label{Nbound}
N\lesssim 5.5+2.1\log_{10}\frac{f_a}{10^{10}\,{\textrm{GeV}}}+2.1\log_{10}\frac{g^{\textrm{max}}_{a\gamma\gamma}}{10^{-10}\,{\textrm{GeV}}^{-1}}.
\ee
%%%
Eq.~(\ref{m0bound}) implies that, unless the couplings $\kappa$ or the fermion Yukawas are tiny, all particles in the spectrum except the axion will be heavy.
If $M>0$, on the other hand, $f$ can in principle be arbitrarily small. In particular one interesting scenario arise if $f$ is around the weak scale, as it may also be motivated by the hierarchy problem. In this case, Eq.~(\ref{m0bound}) can be turned into a lower bound on $M$,
\be\label{mneq0}
M\gtrsim11.2-2.1\log_{10}\frac{f}{100\,{\textrm{GeV}}}-2.1\log_{10}\frac{g^{\textrm{max}}_{a\gamma\gamma}}{10^{-10}\,{\textrm{GeV}}^{-1}}.
\ee
while Eq.~(\ref{Nbound}) now applies to $N-M$. If $f\sim 100$\,GeV and $\kappa\lesssim 1$, the heavy angular modes are around the weak scale. 

These angular modes, a trademark of the clockwork mechanism, are the subject of the next section where we will be discussing their phenomenology in both limiting cases $M=0$ and $f\sim 100$\,GeV.

%%%%%%%%%%%%%%%%%%%%%%%%%%%%%%%%%%%%%%%
\subsection{Phenomenology of clockwork pseudoscalars}
%%%%%%%%%%%%%%%%%%%%%%%%%%%%%%%%%%%%%%%% 
We have identified two minimal renormalizable models defined by the combination of vector-like fermions $\psi_E+\psi_U$ and $\psi_E+\psi_D$ at sites $M$ and $N$ respectively.
The interactions of the angular modes with the SM are inherited from the couplings in Eq.~(\ref{qcdfermions}) and (\ref{photon-fermions}). Neglecting the radial modes they give
\be
\frac{y_M C}{\sqrt 2} fe^{\frac{i\pi_M}{Cf}} \psi_E \psi_{E^c}+\frac{y_N C}{\sqrt 2} f e^{\frac{i\pi_N}{Cf}} \psi_{r} \psi_{r^c},\quad\quad \psi_{r}=\psi_{U}\,\,\mathrm{or}\,\,\psi_{D}.
\ee
As already pointed out the the fermions get a Dirac mass $m_{\psi_E,\psi_{r}}=y_{M,N}C f/\sqrt 2$. Assuming $\kappa\lesssim y_{M,N}$, so that the fermions are heavier than the angular modes, the couplings to the gauge bosons at leading order in $m_A^2/m_{\psi_E,\psi_{r}}^2$, are obtained performing a field redefinition
\be
\psi_E\rightarrow e^{-\frac{i\pi_M}{Cf}} \psi_E,\quad \psi_{r}\rightarrow  e^{-\frac{i\pi_N}{Cf}}\psi_{r}.
\ee
Due to the anomaly associated with such a transformation, and taking into account the quantum numbers of the fermions, the following interactions are generated
\be
-\frac{g_s^2}{32\pi^2} \frac{\pi_N}{Cf}\, G^{A\mu\nu}\tilde G_{\mu\nu}^A-\frac{g'^2}{16\pi^2} \frac{\pi_M+3Y^2 \pi_N}{Cf}\, B^{\mu\nu}\tilde B_{\mu\nu}%-3Y^2\frac{ g'^2}{16\pi^2} \frac{\pi_N}{Cf}\, B^{\mu\nu}\tilde B_{\mu\nu}
\ee
where $Y=-2/3,1/3$ is the hyper-charge of the fermions $U$ and $D$ respectively, and the factor 3 takes into account the color multiplicity.\footnote{For more generic SM representations for the fermions, the anomalous couplings are given by Eq.~(\ref{coup-ksvz}).} Using Eq.~(\ref{rotation}) to go to the mass basis, one finds the following decays width to gauge bosons
\be\label{widths}
\begin{split}
\Gamma(A_i\to gg) &= |a^{(i)}_{N}|^2 \frac{\alpha_s^2}{32\pi^3} \frac{m_{i}^3}{C^2f^2},\\
\Gamma(A_i\to \gamma\gamma) &= |a^{(i)}_{M}+3Y^2a^{(i)}_N|^2 \frac{\alpha^2}{64\pi^3} \frac{m_{i}^3}{C^2f^2},\\
\Gamma(A_i\to Z\gamma) &= |a^{(i)}_{M}+3Y^2a^{(i)}_N|^2 \frac{\alpha^2\tan^2\theta_W}{32\pi^3}  \frac{m_{i}^3}{C^2f^2},\\
\Gamma(A_i\to ZZ) &=  |a^{(i)}_{M}+3Y^2a^{(i)}_N|^2  \frac{\alpha^2\tan^4\theta_W}{64\pi^3} \frac{m_{i}^3}{C^2f^2}.
\end{split}
\ee

As a limiting case it is useful to start discussing the $M=0$ case. In this case $f$, Eq.~(\ref{m0bound}), is large and the coupling to SM highly suppressed. For masses above the GeV scale, the decay rate of the angular modes is dominated by the coupling to gluons. For $M=0$, all mixing angles are proportional to $\sin i \pi/(N+1)$, which means that, roughly, the minimal and maximal lifetimes are attained by $A_{N/2}$ and $A_{1}$ respectively
\be
5\times 10^{-29}\,{\textrm{s}}\left(\frac{10^7\,{\textrm{GeV}}}{f}\right)\frac{N}{\kappa^{3/2}}\lesssim \tau_A\lesssim 2.5\times 10^{-29}\,{\textrm{s}}\left(\frac{10^7\,{\textrm{GeV}}}{f}\right)\frac{N^{3/2}}{\kappa^{3/2}}.
\ee
Such lifetimes are extremely small unless the quartic $\kappa$ is tiny. When $f\sim 10^7$\,GeV lifetimes as large as a second can be obtained for $\kappa\sim 10^{-17}$, for which the angular modes weight $O(100\,{\textrm{MeV}})$ and decay to photons. In this regime, BBN and CMB constraints~\cite{BBN} demand the pseudoscalars to be short-lived $\tau \lesssim 0.1\ \mathrm{s}$. Moreover, for masses up to  $300$\,MeV, shorter life-times are also constrained by the duration of the neutrino pulse from SN1987A~\cite{SNold}. We will not discuss the small $\kappa$ region any further, even though the presence of many light weakly coupled scalars could have interesting phenomenological implications.  
%In this regime, BBN constraints demand the pseudoscalars to be short-lived $\tau \lesssim 0.1\ \mathrm{s}$, (for this particular mass region, shorter life-times are also constrained by SN1987A). While if the pseudoscalars have masses of $O$(GeV), BBN constraints are avoided.\AT{check?}

%%%%%%%%%%%%%%%%%%%%
%%%%%%%%%%%%%%%%%%%%
%%%%%%%%%%%%%%%%%%%%
\begin{figure*}[tb]
\centering
\includegraphics[width=0.6\linewidth]{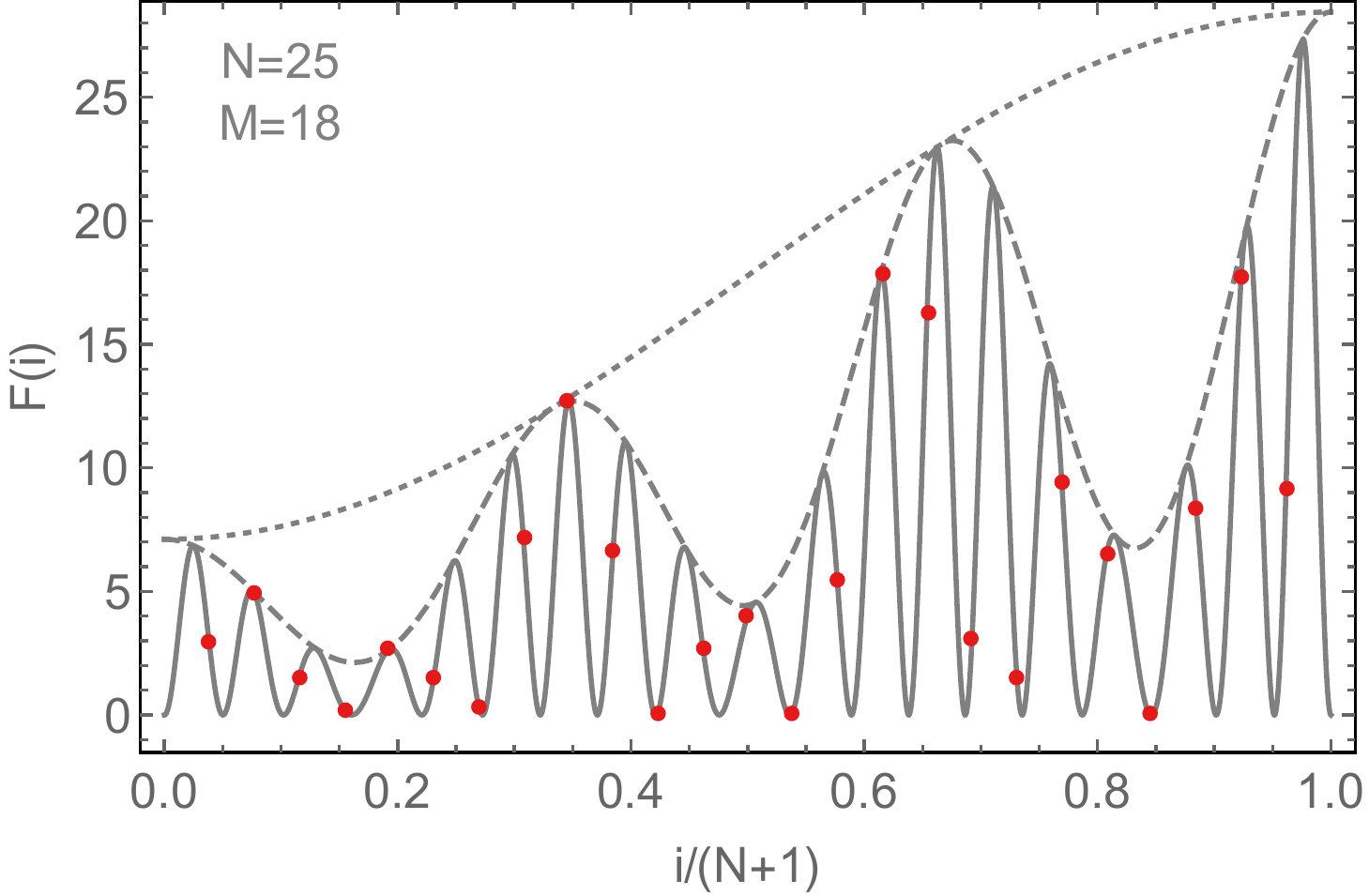}
\caption{{\small\it 
Values of $\mathcal F$ for one special choice of $N$, $M$, and for $Y=1/3$ and $q=3$. For fixed $N$ and $M$, the value of the function $\mathcal F(i)$ lie on the gray curve. Notice that while the specific value of $\mathcal F(i)$ depends on $N$, $M$, and $i$, the envelope of it as a function of $x\equiv i/(N+1)$ only depends on the difference $N-M$ as $\mathcal F_{\textrm{env}}=(1+q^2-2q\cos x\pi)(1+9Y^4+6Y^2\cos(N-M) x\pi)$, which is the gray dashed curve. Finally the upper bound on $\mathcal F$ is given by the universal function $\bar{\mathcal F}(x)$ in Eq.~(\ref{envelope}), which we plot as the gray dotted curve.}}
\label{figF} 
\end{figure*}
%%%%%%%%%%%%%%%%%%%%
%%%%%%%%%%%%%%%%%%%%
%%%%%%%%%%%%%%%%%%%%

Moving now to discuss the case $M\neq0$, Eq.~(\ref{mneq0}) shows that unless $M$ is largish, $O(10)$, $f$ will be much larger than the weak scale and the angular modes very weakly coupled to the SM. It is only when $M\gtrsim 10$ that $f$ is allowed to be small. There is thus an interesting limiting case to consider which is realized when $f\sim 100$\,GeV and $\kappa\lesssim 1$, so that all pseudoscalars have weak scale masses and sizable coupling to the SM and can in principle be produced at the LHC.
As the angular modes couples to gluons and photons the simplest channel to consider is $pp\to A_i\to \gamma\gamma$, through gluon fusion.\footnote{We believe that enough time has passed since Those Days \cite{strumia750} that we should not be held in contempt for considering such a signature.} In the narrow width limit, the cross section for the signal is given by
\be\label{sigma}
\sigma(pp\to A\to \gamma\gamma)=\frac{C_{gg}(m_A^2/s)}{s}\times \frac{\Gamma_{A\to gg}{\Gamma_{A\to \gamma\gamma}}}{m_A\Gamma_A}
\ee
where $s$ is the collider center of mass energy, and $C_{gg}$ is a dimensionless integral of the gluon parton distribution
\be\label{cggeq}
C_{gg}(\tau)=\frac{\pi^2}{8}\int_{\tau}^1\frac{dx}{x}f_g(x) f_g\left(\frac{\tau}{x}\right).
\ee
%The function $C_{gg}$ is plotted in the left panel of Fig.~\ref{cgg} as a function of $m_A$ for $\sqrt{s}=8,13$\,TeV, using NNPDF2.3@NNLO~\cite{nnpdf} with $\alpha_s=0.119$. 
To simplify further the discussion, we notice that for all the massive pseudoscalars $\Gamma_A$ is typically dominated by the two gluons channel. In this case we can write more explicitly
\be
\sigma(pp\to A\to \gamma\gamma)\approx \frac{C_{gg}(m_A^2/s)}{s}\times \frac{\Gamma_{A\to \gamma\gamma}}{m_A}.
\ee
Using Eq.~(\ref{widths}) we obtain, for given $N$, $M$ and $i$
\be
 \frac{\Gamma_{A_i\to \gamma\gamma}}{m_{A_i}}=\frac{\alpha^2\kappa}{64\pi^3(N+1)}\times \mathcal F(i),
 \ee
 where 
 \be
  \mathcal F(i)\equiv \left(\sin\frac{M i\pi}{N+1}-q\sin\frac{(M+1)i\pi}{N+1}-3 Y^2 q\sin\frac{Ni\pi}{N+1}\right)^2.
\ee
The function $\mathcal F$ is a quite complicated oscillating function of $i$ displaying beats. We show it in Fig.~\ref{figF} for $q=3$ and $Y=1/3$ and various choices of $N$ and $M$.  We notice however that despite its intricate dependence on $N,M$ and $i$, $\mathcal F$ is strictly bounded from above by
\be\label{envelope}
\bar{\mathcal F}(x)=(1+3Y^2)^2(1+q^2-2q\cos x\pi), \quad x\equiv i/(N+1).
\ee
Notice in particular that $\bar{\mathcal F}$ is just the function $\lambda$ of Eq.~(\ref{massive}) describing the masses of the angular modes, apart from a numerical prefactor. These considerations suggest that we can write an upper bound for the signal of Eq.~(\ref{sigma}) as
\be\label{upperbound}
\sigma(pp\to A_i\to \gamma\gamma)\times \frac{s}{C_{gg}(m_i^2/s)}\leq\frac{\alpha^2(1+3Y^2)^2}{32\pi^3(1+N)}\frac{m_i^2}{C^2f^2}
\ee
where, for fixed $f$ and $q=3$
\be
\sqrt{2\kappa} \leq \frac{m_i}{Cf}\leq 2\sqrt{2\kappa}.
\ee

%%%%%%%%%%%%%%%%%%%%
%%%%%%%%%%%%%%%%%%%%
%%%%%%%%%%%%%%%%%%%%
\begin{figure*}[tb]
\centering
\includegraphics[width=0.65\linewidth]{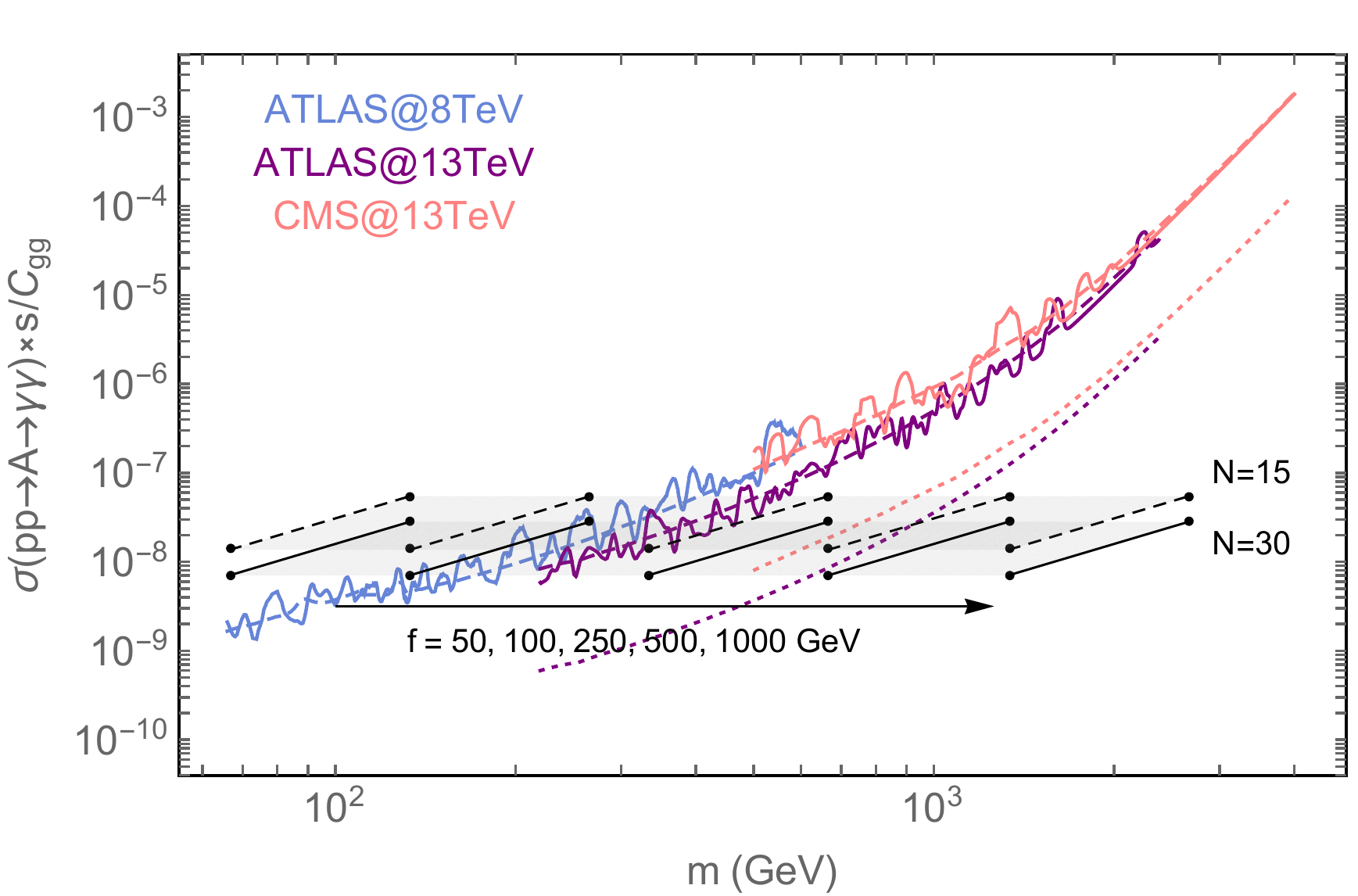}
\caption{{\small\it 
Experimental limits (rescaled by the gluon fusion parton luminosity) for a diphoton resonance. We show observed and expected limits from ATLAS ~\cite{atlas8,atlas13} and CMS \cite{cms13} both at 8 and 13 TeV (solid lines). We also present projections for LHC13 with $3000/\mathrm{fb}$ (dashed lines). The black lines superimposed on the limits show the upper bound, Eq.~(\ref{upperbound}), on the signal cross section predicted by the clockwork QCD axion for $f=50,100,250,500,1000\,{\textrm{GeV}}$ from left to right and two choices of $N$, $N=15$ (dashed) and $N=30$ (full).}}
\label{cgg} 
\end{figure*}
%%%%%%%%%%%%%%%%%%%%
%%%%%%%%%%%%%%%%%%%%
%%%%%%%%%%%%%%%%%%%%

While a precise bound on a specific clockwork model, depending on $N$, $M$, $f$, and $\kappa$, would be quite difficult to obtain, Eq.~(\ref{upperbound}) allows to understand very easily how small $f$ can be and still be unconstrained by the LHC. In Fig.~\ref{cgg} we draw the existing upper bounds on $\sigma\times s/C_{gg}$, which is the relevant dimensionless quantity, and compare them with the upper bound in Eq.~(\ref{upperbound}), fixing $\kappa=1$ and $Y=1/3$, for various choices of $f$ and two values of $N$. The effect of varying these parameters is very easy to understand from Eq.~(\ref{upperbound}): increasing/decreasing $f$ translates the curve right/left, decreasing/increasing $N$, or in general increasing/decreasing the combination $(1+3Y^2)/\sqrt{N+1}$, translate the bound upward/downward, finally, decreasing/increasing $\kappa$ translate the curve downward/upward parallel to itself.
The plot shows that small values of $f$ are still allowed: $f$ can be as low as 250\,GeV for $N=15$ and as small as 200\,GeV if $N=30$. In the same plot we also show the expected final reach of the LHC obtained by rescaling the current expected limit according to the final integrated luminosity which is assumed to be $3000\,{\textrm{fb}}^{-1}$.

%%%%%%%%%%%%%%%%%%%%
%%%%%%%%%%%%%%%%%%%%
%%%%%%%%%%%%%%%%%%%%
\begin{figure*}[tb]
\centering
\includegraphics[width=0.65\linewidth]{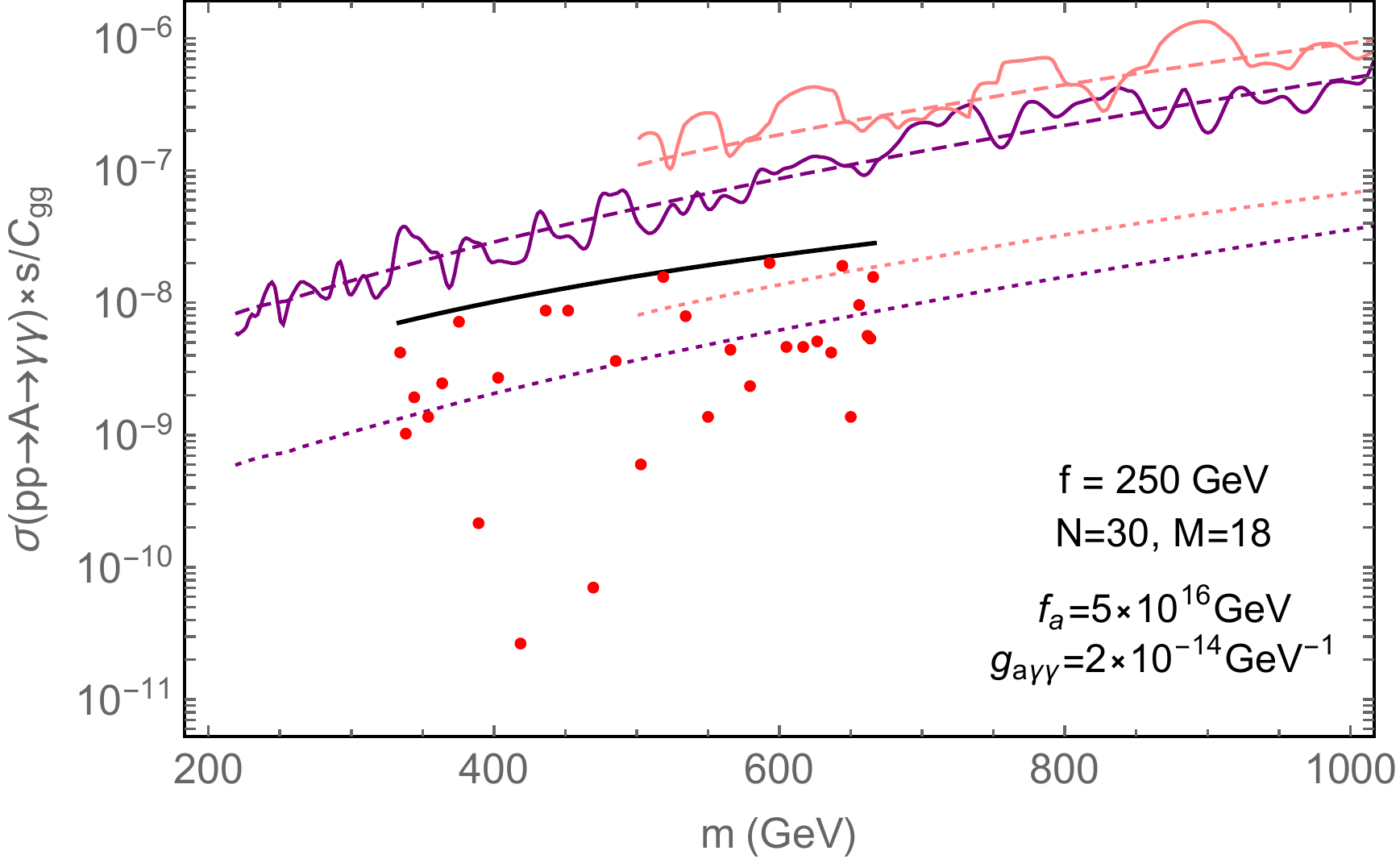}
\caption{{\small\it  For a given choice of parameters in the clockwork QCD axion model (corresponding to the star in Fig.~\ref{plottone}) which allows detection by multiple axion DM detection experiments (ABRACADABRA, Stage-II CASPEr-EDM and CASPEr-wind), we show the expected signal from the angular modes in LHC searches for di-photon resonances (red dots), together with the upper bound in Eq.~(\ref{upperbound}) (black line). $\kappa=1$ and the field content is the minimal one described at the beginning of this section. The LHC limits are the same as in Fig.~\ref{cgg}.}}
\label{figexample} 
\end{figure*}
%%%%%%%%%%%%%%%%%%%%
%%%%%%%%%%%%%%%%%%%%
%%%%%%%%%%%%%%%%%%%%

One of the most important consequences of our construction is the possibility, perhaps overly optimistic, of multiple signals in different experiments.
In Fig.~\ref{figexample} we consider a set of parameters for which axion DM can in principle be detected by both ABRACADABRA and CASPEr (during its second phase of operations). This special point is marked by a yellow star on Fig.~\ref{plottone}. Furthermore since the fundamental axion decay constant $f$ is chosen to be small (but not too small to be already excluded) a striking signal will show up at the LHC, where many resonances decaying to $\gamma\gamma$ (and in principle to $Z\gamma$ and $ZZ$) will be observed in a narrow mass range between $350$ and $650$\,GeV, allowing to reconstruct the parameters of the model. The dijet rate will behave similarly to Eq.~\eqref{upperbound} with the replacement $\alpha\to2\alpha_s/(1+3Y^2)$ for the model under consideration. While below current experimental limits, it is possible that future CMS dijet scouting searches could reach the necessary sensitivity~\cite{scouting}.

\section{Discussion and outlook}
\label{sec:conclusions}

Before we conclude our discussion we would like to say a few words on how to reconcile this scenario with models that offer a solution to the ever lasting Naturalness problem. 

The model has two main fundamental parameters, the scale $f$ and the number of sites $N$, that in 
our discussion have been treated as free parameters. However, as observed in \cite{giudice}, the presence of the scale $f$ can be connected with the inverse size of an extra-dimension with each of the $N$ sites corresponding to a location on such extra-dimension. When the analogy with the continuum limit is pushed even forward, the authors of \cite{giudice} have also realized that the ``clockwork'' mechanism is reproduced by a 5D metric dual to some Little String Theory models \cite{LST2,LST1}. A value of $f$ around the weak scale would solve the hierarchy problem in a way analogous to how large extra dimensions solve it: the ultimate cutoff of the theory, the scale of quantum gravity, is low, around the weak scale. We will not comment further about this point of view but we notice that it may be possible to realize a photo-philic QCD axion with weak scale $f$ in this setup, by introducing an additional brane on which the electromagnetically  charged fermions are localized.

Another possibility to connect our discussion with a conventionally natural framework is offered by supersymmetry and was sketched in~\cite{rattazzi}. Because of holomorphicity, the existence of a PQ symmetry requires the presence of at least a pair of chiral superfields $\Phi_j$ and $\bar{\Phi}_j$ with $j=0,1,...,N$, with a superpontential given by
\be\label{susy}
\mathcal{W}=\epsilon\sum_{j=0}^{N-1} (\Phi_j\bar{\Phi}_{j+1}^2 + \bar\Phi_j \Phi_{j+1}^2) + \mathcal{\widetilde{W}}(\Phi,\bar{\Phi},\cdots),
\ee
where the first term has two unbroken U(1)$_{1,2}$ symmetries with hierarchical charges. In order to realize the clockwork mechanism, $\widetilde{W}$ and possibly the soft terms need to break spontaneously a U(1)$^{N+1}$ symmetry that acts on all the $\Phi,\bar\Phi$ and that includes a linear combination of U(1)$_{1,2}$ as a subgroup. 
After the spontaneous breaking at a scale $f$, and below the scale of the massive fields, the Goldstones can be parametrized by $\Phi_j =f \exp(\Pi_j/f),\, \bar{\Phi}_j = f \exp( -\Pi_j/f)$. Expanding the superpotential we recover the same mass matrix for the $\Pi_i$ as in Eq.~\eqref{masstermCW} but with $q=2$. There is an axion and $N$ massive states with $m_{i}=\epsilon f \lambda^{(i)}$, notice the different scaling with $\lambda$ as compared to the non-supersymmetric case. In case U(1)$^{N+1}$ is mostly broken by soft terms, $f$ would be proportional to the scale of superpartners, rather than being a supersymmetric parameter, opening the connection with supersymmetric scenarios at the TeV or intermediate scale. We think that these interesting theoretical aspects deserve future attention.\\

Irrespective of the UV description, however, we believe that our discussion already captures the phenomenological aspects of the clockwork scenario. The main novelty is a QCD axion with a very large coupling to photons. This situation is typical in ``clockwork'' models, since the effective decay constant of the axion depends exponentially on the `distance' between two sites in the ``clockwork'' chain. Therefore, if the sites with QED and QCD interactions are sufficiently `distant' in theory space, an exponential hierarchy is naturally generated among the coupling to photons and gluons. The interesting case is when the photon coupling is exponentially large, since this opens a new region in the parameter space for the QCD axion. As depicted in Fig.~\ref{plottone} the clockwork can be very similar to the standard QCD axion, but it can also deviate substantially from it. An interesting benchmark is where the axion is DM and in the reach of planned future experiments such as ABRACADABRA and CASPEr, but also ADMX and IAXO. We would like to stress again that this region could not be reached with standard QCD axion models, since it would require exponentially large electric charges $Q$ or an exponentially large number of new particles $N_p$, but perturbativity requires $Q^2N_p\lesssim 4\pi/\alpha$.

In general the parameters $f$ and $N$ are free. However, an interesting correlation between a photo-philic QCD axion and LHC physics arises when the fundamental scale $f$ is around the weak scale. The $N$ pseudoscalars have now masses within the reach of LHC and sizable coupling to photons and gluons. Despite the huge number of fields, the rates can be computed analytically thanks to the exact diagonalization of the mass matrix.
We have found that the most interesting channel is diphoton, where we expect a large number of scalars to have sizable rates. Another important decay channel is $Z\gamma$ where the rate is just a factor of $2\tan^2\theta_W$ smaller than diphoton. 
In Fig.~\ref{cgg} the projected sensitivity of LHC with 3000 fb$^{-1}$ suggests that a relevant part of the parameter space will be explored. In this case we show, as an example, a choice of parameters where we can have signals in ABRACADABRA and CASPEr and, at the same time, the model can be discovered at the LHC.

Finally we remind the reader that we assumed for simplicity the couplings $m$, $\lambda$ and $\kappa$ do not vary from site to site. Furthermore we imposed $\kappa\ll \lambda$ to decouple the physics of the radial modes. The first assumption is not strictly necessary for the clockwork mechanism to operate and deviation from it will affect the LHC phenomenology of the model in a way that can be interesting to study. The radial modes can be, on the other hand, an additional source of interesting LHC signatures that deserve further study.

%We believe that, despite its simplicity, this model can serve as another motivation to explore fully the parameter space of the QCD axion.

%

%%%%%%%%%%%%%%%%%%%%%%%%%%%%%%%%%%%%%%%%%%%%%%%%%%%%%%%%%%%%%%%%
%%%%%%%%%%%%%%%%%%%%%%%%%%%%%%%%%%%%%%%%%%%%%%%%%%%%%%%%%%%%%%%%%%%%%%%%%%%%%%%%%%%%%%%%%%%%%%%%%%%%%%%%%%%%%%%%%%%%%%%%%%%%%%%%%%%%%%%%%%%%%%%%%%%%%%%%%%%%%%%%%%%

\subsubsection*{Acknowledgments}
M.F. is supported in part by the DOE Grant DE-SC0010008. D.P. is supported by the NSF CAREER grant PHY-1554858. F.R. is supported by the DFG Graduiertenkolleg GRK 1940 â€``Physics Beyond the Standard Model" and by the Heidelberg Graduate School of Fundamental Physics. The work of A.T. is supported by an Oehme fellowship. Finally, F.R. would like to thank the CCPP-NYU for its hospitality. 
%%%%%%%%%%%%%%%%%%%%%%%%%%%%%%%%%%%%%%%%%%%%%%%%%%%%%%%%%%%%%%%%%%%%%%%%%%%%%%%%%%%%%%%%%%%%%%%%%%%%%%%%%%%%%%%%%%%%%%%%%%%%%%%%%%%%%%%%%%%%%%%%%%%%%%%%%%%%%%%%%%%%%%%%%%%%%%%%%%%%%%%%%%%%%%%%%%%%%%%%%%%%%%%%%%%%%%%%%%%%%%%%%%%%%%%%%%%%%%%%%%%%%%%%%%%%%%%%%%%%%%%%%%%%%%%%%%%%%%%%%%%%%%%%%%%%%%%%

%

%%%%%%%%%%%%%%%%%%%%%%%%%%%%%%%%%%%%%%%%%%%%%%%%%%%%%%%%%%%%%%%%%%%%%%%%%%%%%%%%%%%%%%%%%%%%%%%%%%%%%%%%%%%%%%%%%%%%%%%%%%%%%%%%%%%%%%%%%%%%%%%%%%%%%%%%%%%%%%%%%%%%%%%%%%%%%%%%%%%%%%%%%%%%%%%%%%%%%%%%%%%%%%%%%%%%%%%%%%%%%%%%%%%%%%%%%%%%%%%%%%%%%%%%%%%%%%%%%%%%%%%%%%%%%%%%%%%%%%%%%%%%%%%%%%%%%%%%%%%%%%%%%%%%%%%%%%%%%%%%%%%%%%%%%%%%%%%%%%%%%%%%%%%%%%%%%%%%%%

\pagestyle{plain}
\bibliographystyle{jhep}
\small
\bibliography{biblio}

\end{document}